# 4 states protocol for time coding quantum key distribution

**Thierry Debuisschert, William Boucher**
*THALES Research and Technology, Domaine de Corbeville, 91404 Orsay Cedex, France*
*Thierry.debuisschert@thalesgroup.com*

**Abstract :** Coherent one photon pulses are sent with four possible time delays with respect to a reference. Ambiguity of the photon time detection resulting from pulses overlap combined with interferometric measurement allows for secure key exchange.
© 2003 Optical Society of America

The information is coded on coherent one photon pulses with a square profile, a duration T and a chosen delay with respect to a time reference. The delays are chosen by Alice so that possible pulses may overlap. Bob can perform only one time measurement which may lead to an ambiguity on the delay evaluation. Bob sends at random half of the pulses he receives to an unbalanced Mach-Zender interferometer (propagation time difference T/2 and phase difference of π) (Fig. 1) to ensure that he actually receives pulses of duration T [1]. The other arm of the input beamsplitter is sent to the photon counter which is used to establish the key between Alice and Bob (Fig. 1).

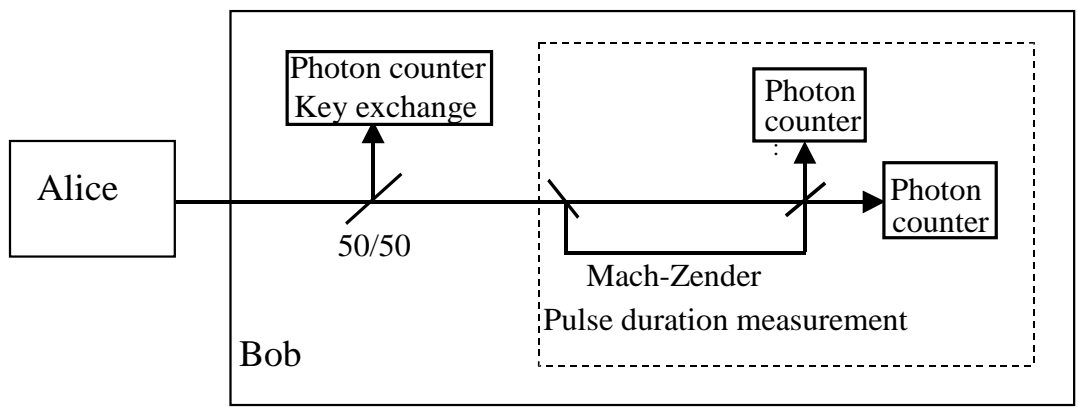

Fig. 1 : Principle of the experiment. Half of the photons received by Bob are used to establish the key. Half are sent to an interferometer for pulse duration measurement

The protocol is modeled with a one photon states $|i\rangle$ basis corresponding to successive pulses of duration T/2. The states used in the protocols are of the form $|i, i+1\rangle = 1/\sqrt{2}\left(|i\rangle + |i+1\rangle\right)$. In the two states protocol [1], Eve can exploit the channel losses to eavesdrop the key. To overcome that limitation, we propose a protocol based on 4 states. Alice sends at random, with equal probabilities, states $|1,2\rangle$, $|2,3\rangle$, $|3,4\rangle$, $|4,5\rangle$ (delays 0, T/2, T, 3T/2 respectively). The information is coded on states $|2,3\rangle$ and $|3,4\rangle$ (eg bit 0 and 1). Detection in time slot 3 do not allow to distinguish between bits 0 and 1. The main difference with the two states protocol is that the useful time slots 2 and 4 are ambiguous during the transmission of the raw key due to the possible emission of states $|1,2\rangle$ and $|4,5\rangle$ by Alice. When Eve measures the pulse position, she has to resend a pulse of duration T, in the ideal case, to avoid a drop of the contrast at Bob interferometer. She then unavoidably induces errors that Alice and Bob can detect comparing publicly part of the raw key and measuring the QBER. After the transmission of the raw key, Alice reveals to Bob each time she has sent states $|1,2\rangle$ or states $|4,5\rangle$ and discard the corresponding measurements. The detection in time slots 2 or 4 becomes non ambiguous to Bob. He then reveals to Alice when this has occurred without revealing the result and discards the ambiguous measurements in time slot 3. A secret key can thus be exchanged between Alice and Bob.

The state that can be resent by Eve can be evaluated as a function of the losses and decoherence induced by the transmission, leading to the maximum allowed probability for Eve to intercept the pulse without being detected. Finally one can calculate the information obtained on Alice by Eve (IAE) and Bob (IAB) respectively [2], as a function of the QBER (Fig. 2). The security criterion IAB > IAE [2] gives a maximum acceptable value of the QBER of 17 % (with no decoherence). One advantage is that there is no need to implement an interferometer between Alice and Bob. One need to synchronize precisely two clocks between Alice and Bob and to keep the coherence of the pulses through the propagation.

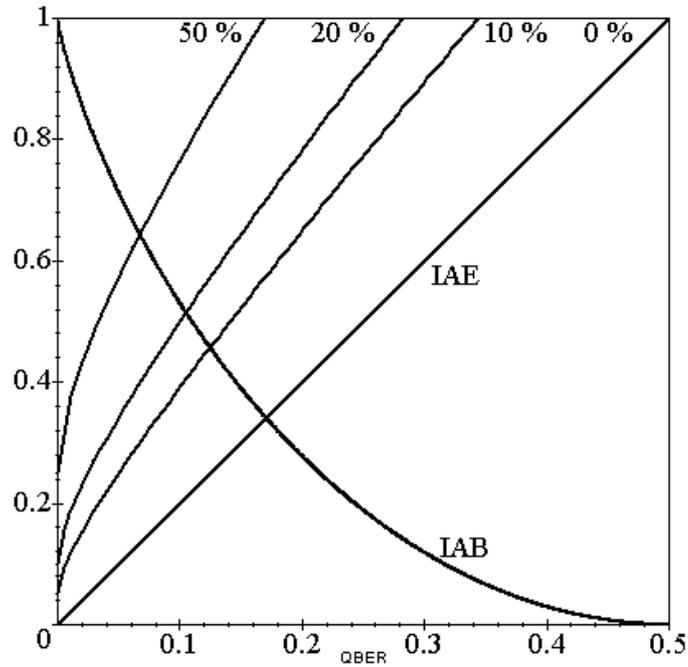

Fig. 2 : Information of Bob on Alice (IAB) and Eve on Alice (IAE) for a decoherence of 0, 10, 20, 50 % of the pulses